\documentclass[12pt]{iopart}

\usepackage{hyperref}
\usepackage{graphicx}
\usepackage{rotating}
\usepackage{array}
\usepackage{amssymb}

\def\v#1{{\bf#1}}
\def\be{\begin{equation}}
\def\ee{\end{equation}}
\def\bea{\begin{eqnarray}}
\def\eea{\end{eqnarray}}

\def\ie{{\it i.e.\,}}
\def\etal{{\it et al.   }}

\def\ecal{\mbox{$\cal E\,$}}

\def\<{\langle}
\def\>{\rangle}

\begin{document}

\title{Time dependent Stark ladders: Exact propagator and caustic control}
\author{E. Sadurn\'i }

\address{Instituto de F\'isica, Benem\'erita Universidad Aut\'onoma de Puebla,
Apartado Postal J-48, 72570 Puebla, M\'exico}

\eads{ \mailto{sadurni@ifuap.buap.mx}}

\begin{abstract}

\noindent
In this note we present a new propagator for a particle in discrete space under the influence of a time-dependent field. With this result we are able to control the shape of caustics emerging from a point-like source, as the explicit form of the wavefronts can be put in terms of the external field. 

\end{abstract}


\maketitle

Similar propagation phenomena in time domain can be found in quantum-mechanical systems, electromagnetic waves and sound waves. The analogies between their wave equations in controlled and well designed situations have allowed the emulation of crystalline structures in settings with periodic symmetry, reaching recently a realization of graphene with microwave cavities \cite{richter}, \cite{sadurni} and photonic crystals \cite{ochiai}. For discrete systems without periodic symmetry, the Stark ladder introduced by Wannier more than fifty years ago \cite{wannier} offers itself as an interesting example. In this respect, we note that the emulation of electrons under the influence of a constant force has been achieved as well: The Wannier-Stark ladder in vibrations of aluminum rods \cite{flores} and the observation of Bloch oscillations in photonic structures \cite{sapienza} seem to be the simplest realizations. Among the most sophisticated, we may single out the propagation of Bose-Einstein condensates in periodic optical traps \cite{sapienza, haensch, morsch} with the possibility of producing a Stark ladder by means of a gravitational field. 

In this note we study the more general case of a homogeneous force field modulated by an arbitrary time-dependent intensity, with the aim of offering another interesting possibility to the already existing configurations and emphasizing the external control of the system through such a field. We shall refer to it as a time-dependent Stark ladder, pointing out to its generalization through the time dependence of the potential and not merely to an equispaced spectrum. Our task is therefore to find the corresponding propagator in closed form. In the case of emulations outside of the quantum regime, we may simply refer to our result as the Green's function. Armed with the result, we shall proceed to characterize the behaviour of caustics emerging from a point-like initial condition. We shall also find the explicit relation between the propagation of the corresponding wavefronts and the time-dependent modulation of the discrete potential, giving the opportunity to discuss the maximal speed of propagation of a signal and how it can be controlled within the restrictions of the Lieb-Robinson bound \cite{lieb}. Two recent studies in theoretical \cite{borovyk} and experimental \cite{cheneau} grounds exemplify the relevance of these ideas.

We start with the problem of finding the propagator for a discrete Schr\"odinger equation in the presence of a time-dependent field. One possible approach for introducing such an equation is by using the central discretization of derivative operators, \ie 

\bea
\fl - \frac{\hbar^2}{2 \mu a^2} \left[ \phi_{n+1}(\tau) + \phi_{n-1}(\tau) - 2 \phi_n(\tau) \right] + a n E(\tau) \phi_{n}(\tau) = i \hbar \frac{\partial \phi_n (\tau)}{ \partial \tau},
\label{s1}
\eea
where $a$ is the lattice spacing, $\mu$ is the mass of the particle and $E(\tau)$ is the external field with dimensions of $($energy$)/($distance$)$. In solid state physics, one may find a similar discrete Schr\"odinger equation arising from a tight-binding model of a single band. Denoting the hopping energy or intersite coupling by $\Delta$ and using the basis of atomic functions $|\phi \> = \sum_n \phi_n | n \>$ , one has
 \bea
\Delta \left[ \phi_{n+1}(\tau) + \phi_{n-1}(\tau) \right] + E(\tau) n \phi_n(\tau) = i \hbar \frac{\partial \phi_n (\tau)}{ \partial \tau},
\label{s2}
\eea
where the field $E(\tau)$ now has the dimensions of energy. Both (\ref{s1}) and (\ref{s2}) can be simplified by means of a convenient redefinition of units and gauge transformations. In this paper we shall work with the tight-binding simplification
\bea
\psi_{n+1}(t) + \psi_{n-1}(t) + \alpha(t) n \psi_{n}(t) = i \frac{ \partial \psi_n (t)}{\partial t},
\label{1}
\eea
which can be obtained from (\ref{s1}) through the following definitions: $t = -\hbar \tau / 2 \mu a^2$, $\alpha(t) = 2 \mu a^3 E(\tau) / \hbar^2$ and $\psi_n(t) = e^{-it/2}\phi_n(\tau)$.

Some years ago, Yellin \cite{yellin} found the propagator of (\ref{1}) for the case $\alpha = \rm{constant}$ by means of the algebraic properties of the hamiltonian. For our problem, the hamiltonian operator reads 

\bea
H(t) = T + T^{\dagger} + \alpha(t) N
\label{2}
\eea
with $T$ a discrete translation operator and $N$ the position operator with integer eigenvalues. The action of these operators on the Hilbert space of atomic (localized) functions is given by

\bea
\left( T \psi \right)_n = \psi_{n+1}, \qquad \left( T^{\dagger} \psi \right)_n = \psi_{n-1}, \qquad \left( N \psi \right)_n = n \psi_{n} 
\label{3}
\eea
and they satisfy the algebra

\bea
\left[ T,T^{\dagger}\right] = 0, \qquad \left[ T,N\right] = T, \qquad \left[ T^{\dagger},N\right] = -T^{\dagger}.
\label{4}
\eea
Now it is evident that for a general function $\alpha(t)$ the hamiltonians at different times do not commute:

\bea
\left[ H(t), H(t') \right] = \left[ \alpha(t) - \alpha(t') \right] \left[ T^{\dagger} - T \right].
\label{5}
\eea
In order to obtain the evolution operator corresponding to (\ref{1}), the use of a Dyson series seems mandatory. We may circumvent such a cumbersome calculation by writing down the Mello-Moshinsky (MM) equations for the {\it discrete representation\ }of the evolution operator $U_{n,m}$ (see chapter VII of \cite{mosh}). Such an operator is related to the discrete propagator by the relation $K_{n,m} = \theta(t) U_{n,m}$. Then, we may obtain $K$ by solving the MM equations through elementary techniques for recursion relations.

It is worth to mention that this procedure was used long ago by Moshinsky and Quesne \cite{quesne} with the purpose to show that linear canonical transformations were represented by gaussian kernels. The harmonic oscillator with a time-dependent frequency is a good example of this, as it generates a linear canonical evolution and its propagator is given by a gaussian in the spatial variables \cite{3.2}. Here we employ these tools to show that the corresponding discrete version leads quite naturally to Bessel functions of field-dependent arguments.

First, we solve the equations of motion for the operators $T, T^{\dagger}$ and $N$ in the Heisenberg picture:

\bea
\dot T = -i \alpha(t) T, \qquad \dot T^{\dagger} = i\alpha(t) T^{\dagger}, \qquad \dot N = i \left(T-T^{\dagger}\right).
\label{6}
\eea
Using the convenient definitions

\bea
f(t) = \int_{0}^{t} d\tau \alpha(\tau), \qquad F(t) = \int_{0}^{t} d\tau \exp \left[-i f(\tau)\right],
\label{7}
\eea
we have the following linear evolution map

\bea
T(t) = U^{\dagger} T(0) U = e^{-if} T(0),
\label{8}
\eea
\bea
T^{\dagger}(t) = U^{\dagger} T^{\dagger}(0) U = e^{if} T^{\dagger}(0),
\label{9}
\eea
\bea
N(t) = U^{\dagger} N(0) U = N(0) + iF T(0) - i F^{*} T^{\dagger}(0).
\label{10}
\eea
Only (\ref{8}) and (\ref{10}) are independent equations, but we include (\ref{9}) as it shall turn useful. We note that $N(0), T(0)$ are not canonically conjugate opertators, but this is not a true obstacle, as they are independent variables and we may employ them in the computation of $U_{n,m}$. We might equally resort to the canonical pair

\bea
P \equiv \frac{1}{2i} \left( T-T^{\dagger}\right), \qquad X \equiv  \left\{ \left( T+T^{\dagger} \right)^{-1}, N  \right\}
\label{11}
\eea
evolving under $U$, but using (\ref{8}), (\ref{9}) and \ref{10}) leads to MM equations which are simpler to solve. The sought MM equations are the following recurrence relations:

\bea
\left[ T(0) U \right]_{m,n} = U_{m+1,n} = e^{-if} \left[ U T(0) \right]_{m,n} = e^{-if} U_{m,n-1} 
\label{12}
\eea
\bea
\left[ T^{\dagger}(0) U \right]_{m,n} = U_{m-1,n} = e^{if} \left[ U T^{\dagger}(0) \right]_{m,n} = e^{if} U_{m,n+1} 
\label{13}
\eea
\bea
\left[ N(0) U \right]_{m,n} = m U_{m,n} = \left[ U N(0) +iF U T(0) -iF^* U T^{\dagger}(0)  \right]_{m,n}.
\label{14}
\eea
The first two relations can be easily solved by noting that

\bea
U_{m,n} = e^{-if} U_{m-1,n-1} \Rightarrow U_{m,n} = e^{-i(n+m)f/2} V_{m-n},
\label{15}
\eea
where $V$ is so far an arbitrary function. Replacing (\ref{15}) in (\ref{14}) yields

\bea
(m-n) V_{m-n} = iF e^{if/2} V_{m-n+1} - iF^* e^{-if/2}V_{m-n-1}.
\label{16}
\eea
Finally, we obtain the familiar recursion relation of the Bessel functions \cite{1} by defining $\phi \equiv \mbox{arg} \left( F \right) + \pi/2 + f/2 $, $\rho \equiv |F|$ and $\nu \equiv m-n$. We get 

\bea
\frac{\nu}{\rho} V_{\nu}(\rho) = e^{i\phi} V_{\nu+1}(\rho)+ e^{-i\phi} V_{\nu-1}(\rho),
\label{17}
\eea
which is solved by $V_{\nu}(\rho) = e^{-i\nu \phi} J_{\nu}(2\rho)$, where $J$ is a Bessel function of the first kind. We exclude the Bessel function of the second kind due to its irregular behaviour at $t=0$, violating $U_{m,n}(0)=\delta_{m,n}$. Our new propagator reads

\bea
K_{m,n}(t) = \theta(t) i^{n-m} \left[ \frac{F(t)}{|F(t)|} \right]^{n-m}  e^{-i m f(t)}  J_{m-n}(2|F(t)|),
\label{18}
\eea
where all the functions can be given explicitly in terms of the external field $\alpha$, as indicated in (\ref{7}). Some properties of (\ref{18}) can be noted immediately. For example, when $\alpha = \mbox{constant}$, we recover

\bea
|F(t)| = \frac{\sin \left( \alpha t / 2\right)}{\alpha / 2}, \qquad  \left[ \frac{F(t)}{|F(t)|} \right]^{n-m}  e^{-imf} = e^{-i (n+m)\alpha t/2},
\label{19}
\eea
leading to the usual propagator for the time-independent Stark ladder. Another important limit comes from the discrete equation (\ref{s1}), where one can let $a \rightarrow 0$ and obtain the propagator of a particle in continuous variables. For a detailed derivation, see the Appendix. It is also interesting to note that under translations, the propagator `picks up' a phase in the form

\bea
K_{n+d,m+d}(t) &=& \exp \left( -id \int_{0}^{t} \alpha(\tau) d\tau \right) K_{n,m}(t) \nonumber \\ &=& \exp \left( -i \int_{0}^{t} \int_{0}^{d} \ecal(\tau, \xi) d\tau d\xi \right) K_{n,m}(t),
\label{20}
\eea
and reminds us of the contribution of the field $\ecal$ to the classical action of a particle in continuous space. But the most important feature for the present work is that $|F(t)|$ is not necessarily periodic, opening the possibility of modifying Bloch oscillations by the direct control of $\alpha(t)$.

Our interest now is to characterize the propagation of point-like initial conditions in this kind of scenario. Placing the particle initially at the origin gives the probability distribution at time $t$ as $|K_{m,0}(t)|^2$. It is well understood \cite{5} that even in the free case, the speed of propagation $v$ in a tight-binding chain has an upper bound (in our units $v_{\mbox{\small max}}=1$) and that the spreading of point-like wave packets occurs at a finite velocity. This can be recognized also in our general problem by means of the integral representation \cite{3} of the Bessel function \footnote{This is not a spectral decomposition of the propagator, in contrast with the free case.}:

\bea
|K_{m,n}(t)| = |J_{m-n}(2|F|)| = \frac{1}{2\pi} \Big| \int_{-\pi}^{\pi} dk \exp \left[ i(n-m)k + i2|F| \sin k \right] \Big|.
\label{21}
\eea
From this integral, a fold-type caustic can be extracted. The ray equation obtained by the method of stationary phase is

\bea
n-m + 2|F| \cos k_{n,m} = 0.
\label{22}
\eea
Solving for $k_{n,m}$ and replacing in the integral of (\ref{21}), gives the curves

\bea
\fl (n-m) \arccos \left( \frac{|n-m|}{2|F|}\right) \pm \sqrt{4 |F|^2 - (n-m)^2} = \mbox{constant} + 2q \pi, \quad q \in \v Z.
\label{23}
\eea
For the main caustic, we simply have zero phase: $(n-m) \arccos \left( \frac{|n-m|}{2|F|}\right) \pm \sqrt{4 |F|^2 - (n-m)^2} =0$, with solutions

\bea
|n-m| \pm 2 |F(t)| = 0.
\label{24}
\eea
This brings out the famous light-cones in $1+1$ dimensional space-time and separates the propagation (or slow) region from the tunneling (or fast) region.


 \begin{figure}[!h] \begin{center} \begin{tabular}{cc} \includegraphics[scale=0.5]{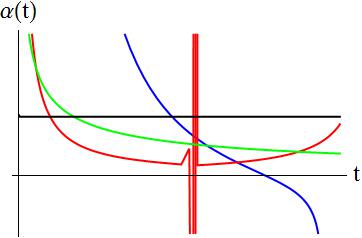} &  \includegraphics[scale=0.5]{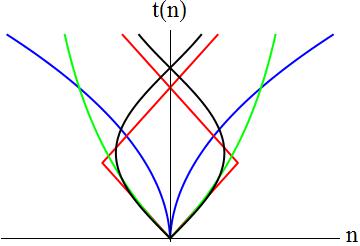} \end{tabular} \end{center} 

\caption{ \label{fig:1} Left panel: Four possible force fields $\alpha(t)$ producing different types motion; the blue curve corresponds to uniform acceleration discussed in example a), the red curve corresponds to a mirror placed at $\rho \neq 0$ in example b), the green curve is the field for an exponential freeze out of the packet expansion in example c) and the constant force in black corresponds to Bloch oscillations. Right panel: The resulting caustics from the previous fields.} \end{figure}

 \begin{figure}[!h] \begin{center} \begin{tabular}{cc} \includegraphics[scale=0.4]{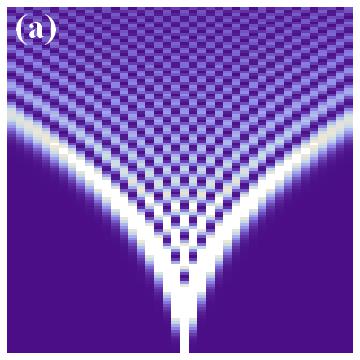} &  \includegraphics[scale=0.4]{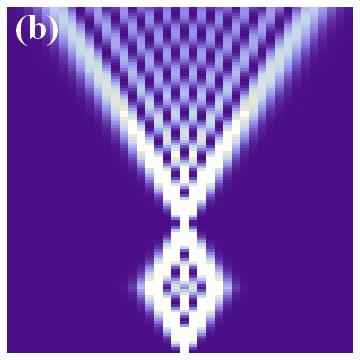} \\  
\includegraphics[scale=0.4]{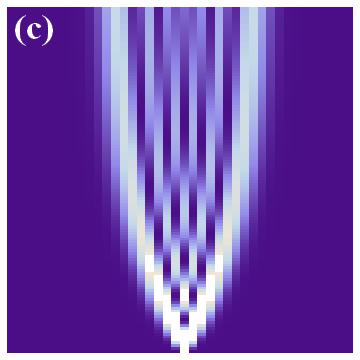} &  \includegraphics[scale=0.4]{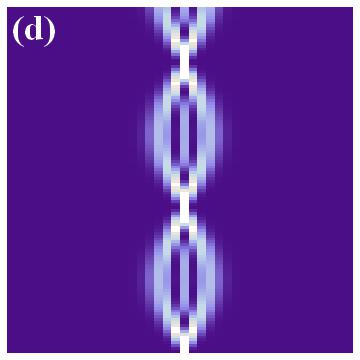} \end{tabular} \end{center} 

\caption{ \label{fig:2} Intensity patterns of a point-like distribution placed at the origin at $t=0$ for different types of external fields. The ordinate represents time and the abscissa represents the {\it discrete\ }coordinate. Panel (a): Uniform acceleration. Panel (b): Mirror at $\rho \neq 0$. Panel (c): Freeze out of the propagation. Panel (d): Bloch oscillations included as a point of comparison.} \end{figure}


Now we turn to the problem of controlling the propagation by means of $\alpha$. We design the shape of our wavefronts in space-time by imposing $|F(t)|$ and solving for the field. This amounts to the inversion of (\ref{7}) and it is a matter of simple algebra to show that

\bea
\alpha(t) = \frac{\sqrt{1-\dot \rho^2}}{\rho} - \frac{\ddot \rho}{\sqrt{1-\dot \rho^2}},
\label{25}
\eea
where $\rho(t) = |F(t)|=\pm |n-m|/2$ gives directly the position as a function of time. The case $\alpha=\mbox{constant}$ can be recovered by setting $\rho$ as a trigonometric function. For propagation speeds $\dot \rho>1$, we note that the second term in (\ref{25}) contains a Lorentz factor that becomes imaginary. Turning $\alpha$ into a complex quantity leads to non-unitary evolution and exponential decrease, in compliance with the Lieb-Robinson bound. In the propagation region, we always have $\dot \rho < 1$. Also notable is the presence of the acceleration of the wavefront $\ddot \rho$ in (\ref{25}), which should not be confused with the acceleration of a classical particle due to the homogenous field. 

Let us demonstrate the use of our formula with three types of `engineered' motion of the wavefronts: a) uniform acceleration, b) uniform velocity with a perfectly reflecting mirror placed at $\rho \neq 0$ and c) exponential `freeze out' of the wave packet expansion.

The condition on $\rho$ for example a) reads $\rho(t)= at^2/2 + v t$, where the intial packet starts at the origin. This results in a field of the form

\bea
\alpha(t) = \frac{\sqrt{1-(at+v)^2}}{ at^2/2 + v t} - \frac{a}{\sqrt{1-(at+v)^2}}.
\label{26}
\eea
The example gives limited motion, since uniform acceleration cannot be sustained forever without obtaining an imaginary Lorentz factor. The motion demands an infinite force at times $t=0,(1-v)/a$ and a vanishing force at an intermediate time. See the blue curves in fig.\ref{fig:1}, where the force and the motion of the wavefront are shown as functions of $t$. The resulting intensity pattern is shown in fig.\ref{fig:2}(a).

For the motion proposed in example b), we set $\rho(t)=1-|vt-1|$ and $v<1$. The field intensity becomes

\bea
\alpha(t) = \frac{\sqrt{1-v^2}}{1-|vt-1|} - \frac{2 \delta(1-vt)}{\sqrt{1-v^2}}.
\label{27}
\eea
Here it is important to note that in order to achieve a uniform $v<1$, we cannot simply turn-off the interaction, as this would strictly produce unit speed of propagation. Instead, we have a non-trivial solution in (\ref{27}) even in the absence of the mirror, producing a slow but uniform motion. At the event of reflection $t=1/v$ (mirror at $\rho=1$) the field has a delta singularity. For a depiction, see the red curves in fig.\ref{fig:1} and panel (b) of fig.\ref{fig:2}.

Finally, example c) provides a method to stop the propagation of the pulse by applying a field. We take $\rho(t) = 1-e^{- \omega t}$, leading to

\bea
\alpha(t) = \frac{ \sqrt{   1-\omega^2 e^{-2\omega t} }  }{   1-e^{-\omega t}   } + \frac{   \omega^2 e^{-\omega t}   }{  \sqrt{   1-\omega^2 e^{-2\omega t} } }.
\label{28}
\eea
The curves for $\alpha$ are shown in green in fig.\ref{fig:1} and the intensity pattern in fig.\ref{fig:2}(c). The resulting force (\ref{28}) tends to a non-zero constant as $t\rightarrow \infty$. However, such a constant field does not produce Bloch oscillations of the packet, since at the origin of time we had a very strong (singular) force and the wavefront depends on the history of the applied field. One can be convinced of this statement by inspecting the time integrals in (\ref{7}).

We conclude this note by emphasizing that exact propagators are rather uncommon objects \cite{3.1}. A collection of these kernels has been given in \cite{3.2} and a number of different paths have been devised for their calculation \cite{3.3}. This has been done mainly for continuous problems, including relativistic ones \cite{sym}. The discrete case should not be an exception. For instance, \cite{5} contains a Feynman path integral version of the free discrete kernel. What we have presented here is a method that is common to both continuous and discrete realms and that explains the resulting solvability for a wide class of systems: The spatial representations of canonical transformations. 

In a less technical order of ideas, let us mention that the examples presented here are of an illustrative character. However, we should not disregard completely their applicability to experiments designed {\it ex professo,\ }which may range from the simplicity of torsional waves in a piece of metal to the ambitious control of Bose-Einstein condensates in a dilute regime.
\pagebreak
\ack

The author is grateful to Rafael M\'endez-S\'anchez and Mois\'es Mart\'inez-Mares for useful discussions, and to Prof. Wolfgang P. Schleich for sharing some of his knowledge on caustics. Financial support from PROMEP Project $103.5/12/4367$ is acknowledged. 

\appendix

\section*{Appendix: Continuous limit }

\setcounter{section}{1}

Here we compute the continuous limit of the propagator corresponding to the Schr\"odinger equation (\ref{s1}). Let us start by writing our kernel in the appropriate units: According to the definitions, we have $\tau = -2 \mu a^2 t / \hbar$, $\psi_n = e^{i \hbar \tau / \mu a^2} \phi_n$ and $\alpha(t) = -2ma^3 E(\tau) / \hbar^2$. This leads to

\bea
\fl K_{nm}(\tau) = i^{n-m} \left[ \frac{F(\tau)}{|F(\tau)|} \right]^{n-m} \exp{\left[-i\left( m f(\tau) + \frac{\hbar \tau}{\mu a^2} \right) \right]} J_{m-n} \left( 2 |F(t)| \right),
\label{ap1}
\eea
where

\bea
f(\tau) = \frac{a}{\hbar} \int_{0}^{\tau} ds E(s),
\label{ap2}
\eea
\bea
F(\tau) = - \frac{\hbar}{2\mu a^2} \int_{0}^{\tau} ds' \exp{\left[ -\frac{ia}{\hbar} \int_{0}^{s'} ds E(s) \right]}.
\label{ap3}
\eea
The continuous limit corresponds to

\bea
a \rightarrow 0, \qquad an \rightarrow x, \qquad am \rightarrow x', \qquad dx \equiv a,
\label{ap4}
\eea
while other quantities such as $\mu, \tau$ and $E(\tau)$ remain fixed. For our computations we need the following ascending expansions in $a$:

\bea
\fl F(\tau) \approx  - \frac{\hbar \tau}{2 \mu a^2} +  \frac{i}{2 \mu a}  \int_{0}^{\tau} ds' \int_{0}^{s'} ds E(s) +  \frac{1}{4 \mu \hbar}  \int_{0}^{\tau} ds' \left[ \int_{0}^{s'} ds E(s) \right]^2
\label{ap5}
\eea

\bea
\fl |F(\tau)| \approx  \frac{\hbar \tau}{2 \mu a^2} +  \frac{1}{4 \mu \hbar \tau}  \left[\int_{0}^{\tau} ds' \int_{0}^{s'} ds E(s) \right]^2 -  \frac{1}{4 \mu \hbar}  \int_{0}^{\tau} ds' \left[ \int_{0}^{s'} ds E(s) \right]^2
\label{ap6}
\eea

\bea
\fl \mbox{arg} \left[ F(\tau) \right] \approx - \frac{a}{\hbar \tau} \int_{0}^{\tau} ds' \int_{0}^{s'} ds E(s).
\label{ap7}
\eea
By substituting (\ref{ap6}) in (\ref{ap1}), we observe that the limit $a\rightarrow 0$ demands the use of an asymptotic form of $J_n(z)$. Such an approximation comes from the Meissel expansion \cite{1} and has the form

\bea
J_n(z) \approx \frac{i^n}{\sqrt{2 \pi i z}} \exp \left[i(z+\frac{n^2}{2z}) \right],
\label{ap8}
\eea
which corresponds to $1 << n << z$. With the expansions (\ref{ap5} - \ref{ap7}) and the asymptotic form (\ref{ap8}), we finally get the limit

\bea
\fl K(x,x';t,0) &=& dx \sqrt{ \frac{\mu}{2 \pi i \hbar \tau} } \exp \left[ \frac{i \mu (x-x')^2}{2 \hbar \tau}\right] 
\exp \left[ \frac{ i (x'-x)}{ \hbar \tau} \int_{0}^{\tau} ds' \int_{0}^{s'} ds E(s) \right] 
\nonumber \\  \fl &\times&  \exp \left\{ -\frac{ i x'}{ \hbar } \int_{0}^{\tau} ds  E(s) -  \frac{i}{2 \mu \hbar } \int_{0}^{\tau} ds' \left[ \int_{0}^{s'} ds E(s) \right]^2 \right\} \nonumber \\ \fl &\times&  \exp \left\{ \frac{i}{2 \mu \hbar \tau } \left[ \int_{0}^{\tau} ds' \int_{0}^{s'} ds E(s) \right]^2 \right\}.
\label{ap9}
\eea
In passing, we note that the resulting asymmetry in $x,x'$ is an effect due to time irreversibility of the external field; the propagator necessarily depends on the two variables $x+x'$ and $x-x'$. One can verify that (\ref{ap9}) is the correct limit of the propagator by checking that it satisfies the equations of motion of the continuous problem. In fact, the MM equations yield 

\bea
\fl \left[ -i \hbar \frac{\partial}{\partial x} -i \hbar \frac{\partial}{\partial x'} \right] K(x,x';t,0) =  \left[ -  \int_{0}^{\tau} ds  E(s)  \right] K(x,x';t,0)
\label{ap10}
\eea
\bea
\fl \left[ x-x' - \frac{i \hbar \tau}{\mu } \frac{\partial}{\partial x} \right] K(x,x';t,0) =  \left[ - \frac{1}{\mu} \int_{0}^{\tau} ds' \int_{0}^{s'} ds  E(s)  \right] K(x,x';t,0)
\label{ap11}
\eea
which are equivalent to the following Heisenberg equations of motion

\bea
p(t) = p_0  - \int_{0}^{\tau} ds E(s)
\label{ap12}
\eea
\bea
x(t) = x_0 + \frac{\tau}{\mu} p_0 - \frac{1}{\mu}  \int_{0}^{\tau} ds' \int_{0}^{s'} ds  E(s).
\label{ap13}
\eea
Furthermore, we can recover the well-known result for a particle in a constant homogeneous field by letting $E(s) = E_0$. We have

\bea
 \fl   -\frac{i E_0^2 \tau^3}{24 \mu \hbar}= \frac{i}{2 \mu \hbar \tau } \left[ \int_{0}^{\tau} ds' \int_{0}^{s'} ds E(s) \right]^2 -  \frac{i}{2 \mu \hbar } \int_{0}^{\tau} ds' \left[ \int_{0}^{s'} ds E(s) \right]^2 
\label{ap14}
\eea
\bea
 \fl -\frac{i (x+x') E_0 \tau}{ 2 \hbar} =\frac{i (x'-x)}{ \hbar \tau} \int_{0}^{\tau} ds' \int_{0}^{s'} ds E(s)  -\frac{ i x'}{ \hbar } \int_{0}^{\tau} ds  E(s)  
\label{ap15}
\eea
and with this result, the phases in (\ref{ap9}) reduce to the ones reported in \cite{3.2}, formula (6.2.18), page 175.

\end{document}